\documentstyle[aps,prl,twocolumn]{revtex}

\begin{document}
\draft


\title{Novel Properties of The Apparent Metal-Insulator Transition in 
Two-Dimensional Systems}

\author{Y. Hanein$^{1,2}$, D. Shahar$^{1,2}$, J. 
Yoon$^{2}$, C.C. Li$^{2}$, D.C. Tsui$^{2}$ and Hadas Shtrikman$^{1}$}

\address{$^{1}$ Dept. of Condensed Matter Physics, Weizmann Institute, 
Rehovot 76100, Israel}

\address{$^{2}$ Dept. of Electrical Engineering, Princeton University, 
Princeton, New Jersey 08544}
\maketitle

\begin{abstract}

The low-temperature conductivity of low-density,
high-mobility, two-dimensional hole systems in GaAs was studied.
We explicitly show that the metal-insulator transition, 
observed in these systems, is characterized by a well-defined 
critical density, $p_{0}^{c}$.
We also observe that the low-temperature conductivity of these 
systems depends linearly on 
the hole density, over a wide density range.
The high-density linear conductivity extrapolates to zero at a density close 
to the critical density.
\end{abstract}


The development of the scaling theory 
of localization \cite{Abrahams1979} contributed a great deal to the 
understanding of the transport properties of electron systems.
For the case of two-dimensional (2D) systems this theory, which neglects 
carrier-carrier interactions, predicts that the electrons are always 
localized at zero temperature ($T$).
This picture stood until, in 1994, Kravchenko 
{\it et al.} \cite{SVKrav94} presented 
experimental evidence supporting the existence of a 
metal-insulator transition (MIT) in 2D. 
The presence of strong electron-electron interactions
\cite{SFinkel84,VDobrosavljevic97,DBelitz97,PPhillips98} was 
proposed to account for the discrepancy between these results and the 
non-interacting theory\cite{Abrahams1979}.
This discrepancy implies the possible revival of other concepts 
that were abandoned following the successes of the  scaling theory 
of localization.
One such  concept is that of minimum metallic conductivity, introduced by Mott in 
1972 \cite{Mott1972}, suggesting that metallic systems can only have a conductivity 
($\sigma$) higher than a minimal value, $\sigma>\sigma_{min}$.
By exploring, in this communication, the low-$T$ conductivity 
characteristics of the apparent
MIT in 2D, we show that it is consistent with the 
existence of a minimum metallic conductivity.
In several samples, $\sigma_{min}$ is close to $e^{2}/h$.

At the absolute zero of temperature the fingerprint of an insulator is a diverging resistivity ($\rho$) 
while that of a metal is a $\rho$ that
reaches a constant value. At finite $T$'s however, all materials (except superconductors)
have a finite $\rho$. 
Consequently, in order to distinguish between insulating and 
metallic phases in experimental situations, 
one has to rely primarily on the extrapolation of the measured $\rho$ 
to $T=0$. 
The temperature coefficient of 
resistivity (TCR), defined as TCR=$\frac{d\rho}{dT}$, can be used to 
determine whether the extrapolated value of $\rho$ 
is likely to be infinite or finite,
and therefore the sign of the TCR is commonly taken to differentiate between insulating and 
metallic phases.
Namely, we label a state as insulator if it has a negative TCR ($\rho$ appears to 
diverge as $T\rightarrow 0$) and a metal if it has a positive, or zero TCR
($\rho$ extrapolates to a finite value at $T=0$).
A change of the TCR from positive to negative as the density 
was lowered was observed by all researchers of 
Ref.\cite{SVKrav94,Pudalov1997,DPopovic97,Colridge,Hanein1998,Simmons1998,SPapadakis1998} for 
various 2D systems, and
was considered indicative of a MIT. 
Although simple and convenient, this procedure is not unambiguous,
and we shall stress the difficulties associated with it below.

Attempts to characterize the new metallic state revealed \cite{Hanein1998,Pudalov1997}
that the metallic-like resistivity of high-mobility 
two-dimensional hole system (2DHS) in GaAs, and of electrons in Si
MOSFET's, is dominated by an 
exponential $T$-dependence and follows
\begin{equation}
\rho(T)=\rho_{0}+\rho_{1}\exp \left( -\frac{T_{0}}{T}\right)
\label{3}
\end{equation}
with a characteristic temperature $T_{0}$, 
which is proportional to the hole-density ($p$). At very low $T$ 
($T<<T_{0}$), where 
the exponential term of Eq. \ref{3} becomes negligible, $\rho(T)$ saturates.
While the exponential term of $\rho$ is responsible for the dramatic 
nature of $\rho$ 
we will focus in this letter on the saturation value, $\rho_{0}$ (or $\sigma_{0}=1/\rho_{0})$, which appears
to be more relevant to the ultimate low-$T$ 
behavior. 
Instead of analyzing $\rho$, we chose to use $\sigma=1/\rho$. They 
are, of course, equivalent but the use of $\sigma$ allowed for an 
observation of a surprising linear dependence of $\sigma$ on $p$,
which is the main finding of this work.

We begin the presentation of our results by pointing out the existence 
of a MIT in one of our high mobility ($\mu$), $p$-type, 
inverted semiconductor insulator semiconductor (ISIS) sample 
\cite{Meirav1988,Hanein1997}. 
Measurements were done in a dilution refrigerator with 
base $T$ of $35$ mK using a standard lock-in technique.
Instead of plotting  $\rho$ versus $T$ 
at various $p$'s we plot, in Fig. 1, isotherms of $\sigma$ versus $p$, 
at several $T$'s between 57 and 840 mK. 
At the high-$p$ range, $\sigma$ decreases 
with $T$ corresponding to a positive TCR. 
At the low-$p$ end, seen in detail in the inset of Fig. 1, $\sigma$ increases 
with $T$, and the TCR is negative .
In the spirit of previous studies, this sign change of TCR identifies
a crossover from an insulating behavior at low-density, to a 
metallic-like behavior at high-density.

Having established the MIT in our system, we now address the low-$T$ behavior of $\sigma$. As can be seen in 
Fig. 1, above 
$\sigma \cong 4 e^{2}/h$, the lower $T$ traces overlap indicating the saturation of 
$\sigma$ at low $T$'s. Therefore for $\sigma \geq  4 e^{2}/h$, we can  regard 
the 57 mK trace as a good representation of $\sigma_{0}$.
Focusing on this curve, a linear dependence of $\sigma_{0}$ on $p$ 
is observed  and $\sigma_{0}$ can be described by (dashed line)
\begin{equation}
	\sigma_{0}=\alpha(p-p_{0}^{e})
	\label{4}
\end{equation}
where $\alpha$ is the linear slope and $p_{0}^{e}$ is the density 
in which the linear fit of $\sigma_{0}$ extrapolates to zero.
This description breaks down for $\sigma  <  3 e^{2}/h$, as we approach 
the transition region.
For a set of samples with mobilities varying 
between $26,000$ and $220,000$ cm$^{2}$/V$\cdot$sec, $\alpha$
changes from $1.8\cdot$ $10^{-10}$ to 
$11\cdot$ $10^{-10}$ 
$e^{2}/h$ cm$^{2}$.

To establish the generality of the linear $\sigma_{0}(p)$ result we show, in 
Fig. 2, similar data for other 2DHS samples.
In Fig. 2a and 2b we show $\sigma$ versus $p$ of a 
high and low-$\mu$ ISIS structures, respectively, and 
in Fig. 2c we depict similar data obtained from a 
2DHS formed in a 10 nm symmetrically doped quantum well.
Indeed, all these samples exhibit a linear $\sigma_{0}(p)$ 
dependence, which holds to high values of $\sigma$.
It is worthwhile 
mentioning that $p$ in our samples is obtained directly from a 
measurement of the Hall effect and is linear with the applied gate 
bias, consistent with a capacitively induced charge-transfer.

The linear $\sigma$ by itself may not have been too surprising, as it 
is a natural result of the Drude model, $\sigma=ne^{2}\tau/m^{*}$, where $n$ is 
the carrier density, $e$ is the electron charge, $\tau$ is the 
elastic scattering time and $m^{*}$ is the carrier effective mass.
However, in our data, $\sigma_{0}$ (see Eq. \ref{4}) does not 
extrapolate to zero at 
$p=0$, but has an offset, $p_{0}^{e}$, of unclear origin.
We will show below that $p_{0}^{e}$ is equal, within 
experimental uncertainty, to the critical point of the MIT.

To determine the critical point of the MIT, we use the vanishing-TCR 
criterion discussed before. Inspecting the 
inset of Fig. 1, we see that each two consecutive isotherms of 
$\sigma(p)$  cross each other at some value of $p$ ($p^{c}$). 
Since at these points the TCR=0, a natural tendency would be to identify 
them as transition points between 
the metallic and the insulating phases.
Accepting this point of view leads to a possible conflict as these TCR=0 
points are clearly $T$ dependent, and an unambiguous 
determination of the transition is therefore impossible.
Fortunately, at the lower $T$ range typically below 150 mK, the 
crossing point appears to be  independent of $T$.  This can be seen 
more readily in Fig. 3, where only low-$T$ traces from Fig. 1 are presented. 
A $T$-independent crossing point, marked by an arrow in Fig. 3,
emerges at a well-defined $p$ 
($p_{0}^{c}$), which clearly separates the conducting and 
insulating phases.
Another way of seeing the settling of $p^{c}$
is by plotting it as function of averaged $T$ 
of each two consecutive curves, as we do in the inset of Fig. 3.
Here, $p^{c}(T)$ decreases monotonically with $T$ for $T$'s above 150 
mK, but for lower $T$'s $p^{c}$ appears to saturate 
to its final value $p_{0}^{c}$.

Based on the results presented above we
can safely argue the following three points. First, at low 
enough $p$, $p<<p_{0}^{c}$,  the system is 
insulating with a vanishingly small $\sigma_{0}$.
Second, for $p>p_{0}^{c}$, metallic-like $\sigma$ is observed down to 
our lowest $T$.
And third, a fixed transition-point 
does exist, and its $p$-value can be identified. 
Having identified a low-$T$ transition point, it is tempting to 
associate it with a MIT. But before we explore the interesting 
consequences of this association we wish to alert the reader to a 
serious caveat that must be born in mind:
For $p$ values close to $p_{0}^{c}$ the saturation is not as clear and 
an unambiguous determination of the phases is impossible.
This is an unavoidable difficulty common to all phase transitions, and 
is not particular to the transition at hand.
Further, for the MIT in three-dimensions (3D), the sign of the TCR is 
certainly not a good indication of the phases, and metallic samples 
can exhibit negative TCR \cite{Dodson1981,Maliepaard}. Nevertheless, 
it is still possible that the 2D MIT will be different in that respect.
With these reservations kept in mind we will proceed with our discussion assuming that 
the low-$T$ fixed point is indeed the MIT.

The first interesting question is the $\rho$ value of the transition 
point, $\rho_{c}$, and whether it is universal. 
For the three different samples for which we were able to reliably extract 
$\rho_{c}$,
it occurs at 
$\sigma_{0}$ $0.83$, $0.9$,  
and 1.58 $e^{2}/h$, which is close to the Yoffe Regel criterion, 
$k_{F}\cdot l$=1. 
This result is similar to the universality found in the 
diagonal $\rho$ for the quantum Hall liquid to Hall insulator 
transition 
\cite{DShahar95}.
This result as well as the observation of a fixed transition point may 
indicate the existence of minimum metallic conductivity in 2D.

If we now focus on the extrapolation of the linear fit of $\sigma_{0}$ 
to $\sigma_{0}=0$, 
denoted in Fig. 3 by a dashed line, we can clearly see 
that $p_{0}^{e}$ 
is close to $p_{0}^{c}$, the low-$T$ crossing point.  
The fascinating consequence of this result is that the 
MIT can be identified by an 
extrapolation from $\sigma_{0}>40 e^{2}/h$, 
much larger than $\sigma_{0}$ at the transition itself.
To further characterize $p_{0}^{e}$ (recall that 
$p_{0}^{e} \approx p_{0}^{c}$), 
we plot, in Fig. 4, $p_{0}^{e}$
for different samples as a function of their mobilities
at $p=5\cdot 10^{10}$ cm$^{-2}$. 
While $\mu$ changes from 220,000 to 26,000 cm$^{2}$/V$\cdot$sec, 
the transition point changes from 0.67 to 2$ \cdot 10 $ cm $^{-2}$.
Namely, higher mobility samples remain conducting 
until lower values of $p$. 
An important consequence of this result lies in the relation between 
the density and the effective strength of the carrier-carrier 
interaction $r_{s}$ ($r_{s} \equiv \frac{U}{E_{F}} 
\propto{\frac{1}{\sqrt{p}}}$). 
The dependence of $p_{0}^{e}$ on $\mu$ indicates that, while the 
strong interactions in these samples are important to the observation 
of the transition, the disorder still plays a significant role, at 
least in determining the $r_{s}$ value where the transition takes 
place.

The $\sigma_{0}$ of Eq. \ref{4} is suggestive of a 
two-component model where a portion of the carriers, $p_{0}^{e}$, contributes only to 
the Hall voltage and not to the longitudinal resistivity.
These carriers may be in a Hall insulator state which is 
characterized by a vanishing $\sigma$ and a Hall resistivity close to 
its classical value.
We note that all samples used in this study are characterized by very large 
separation ($>150$ nm) between the conducting channel and any intentional doping.
This large separation minimizes the scattering from 
ionized dopants \cite{Ando1982}.

To summarize, we used various samples to investigate the density dependence of 
$\sigma$ of 2DHS's in GaAs.   
We provided evidence that, at low-$T$, $\sigma$ changes from 
metallic-like to 
insulating at a well-defined $p$ value, ($p_{0}^{c}$).
If identified correctly as the MIT, 
$p_{0}^{c}$ may describe a point of minimum metallic 
conductivity with a value close to $e^{2}/h$.
We see that for all of our samples, above a critical value of density, 
the low-$T$ conductivity has a linear 
dependence on density. 
The density 
in which the linear fit of $\sigma_{0}$ extrapolates to zero is 
finite ($p_{0}^{e}$).
We show that the value of $p_{0}^{e}$  scales with the
high-$p$ mobility of the 2D system.
We find that  $p_{0}^{e} \sim p_{0}^{c}$.
These results suggest that the transition as well as the physics 
near it are related to the physics far away from the 
transition.

This work was supported by the NSF and by a grant from the Israeli 
Ministry of Science and The Arts.

\begin{figure}  
\caption{ 
$\sigma$ versus $p$ of a high-$\mu$ $p$-type ISIS 
sample at various fixed $T$'s, $T$=57, 69, 71, 
104, 138, 226,
308, 405, 571, 740, 839 mK. 
The dashed line is the fit of the 57 mK data to $\sigma=\alpha 
(p-p_{0}^{e})$.
The inset shows a focus on the low density range of the data shown in Fig. 1.} 
\end{figure}

\begin{figure}  
\caption{
(a) $\sigma$ versus $p$ of a high-$\mu$ ($\mu$=134,000 
cm$^{2}$/V$\cdot$sec at $p=5\cdot 10^{10}$ cm$^{-2}$),
$p$-type ISIS at $T$=50 mK.
(b) $\sigma$ versus $p$ of low-$\mu$ ($\mu$=26,000 
cm$^{2}$/V$\cdot$sec at $p=5\cdot 10^{10}$ cm$^{-2}$), 
$p$-type ISIS at $T$=48 mK.  
(c) $\sigma$ of 2DHS formed in a 10 nm thick symmetrically doped 
quantum well versus $p$ at $T$=35 mK 
($\mu$=65,000 
cm$^{2}$/V$\cdot$sec at $p=5\cdot 10^{10}$ cm$^{-2}$).}
\end{figure}

\begin{figure}  
\caption{Focus on the low density range of the data shown in Fig. 1, 
only low-$T$'s are shown,  $T$= 57, 69, 71, 
104, 138 mK.  The dashed line is the extrapolation of the fit 
of the high-$p$ data to $\sigma=\alpha (p-p_{0}^{e})$.
The arrow marks the low-$T$ crossing point, $p_{0}^{c}$.
The inset shows the crossing point between each two consecutive isotherms
of $\sigma(p)$, plotted versus the averaged 
$T$ of each two curves.}
\end{figure}

\begin{figure}  
\caption{The critical density, $p_{0}^{e}$, 
versus $\mu$ (at $p=5\cdot10^{10}$ cm$^{-2}$) for various 
2DHS samples. $p_{0}^{e}$ appears to decrease monotonically with $\mu$.}
\end{figure}

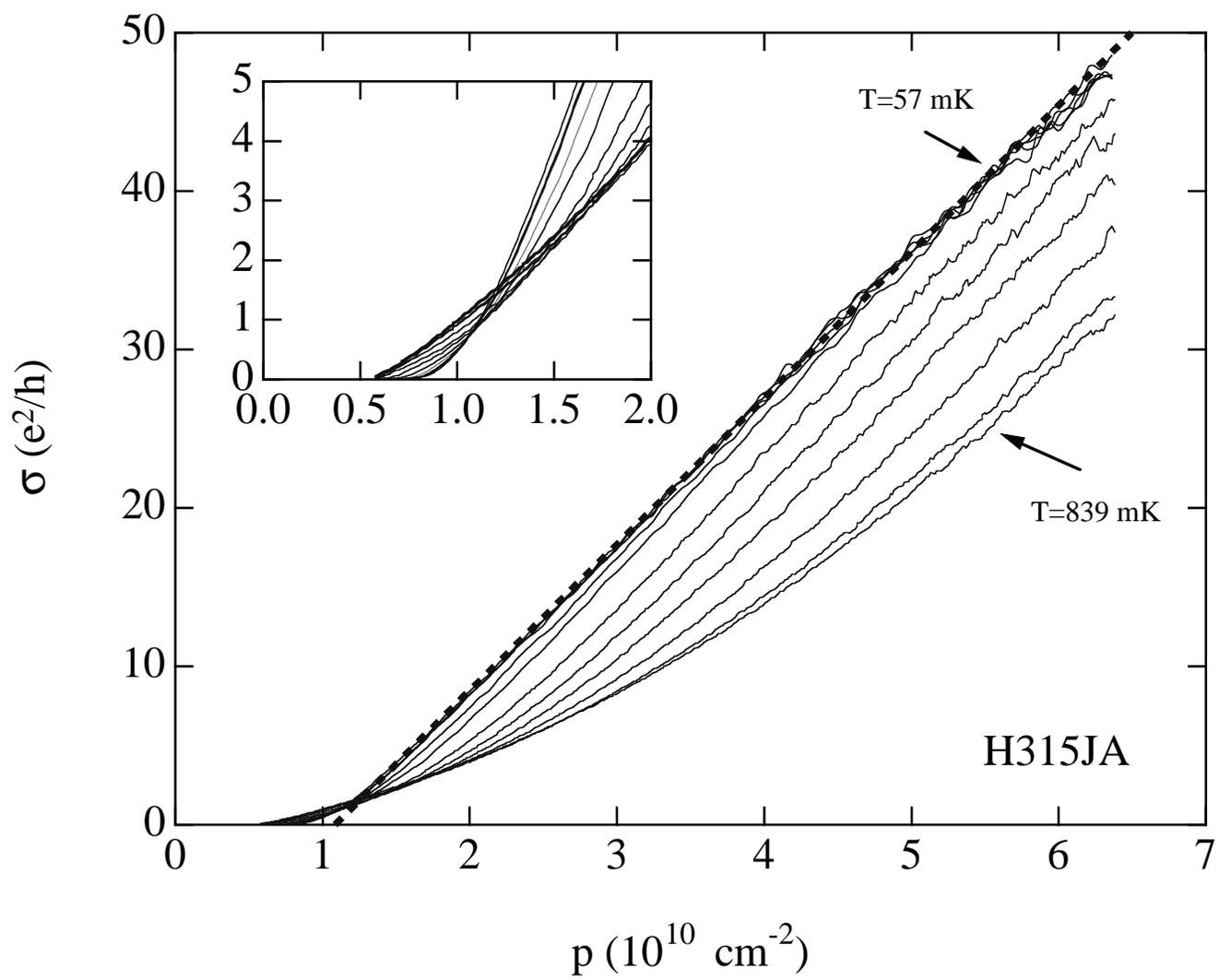
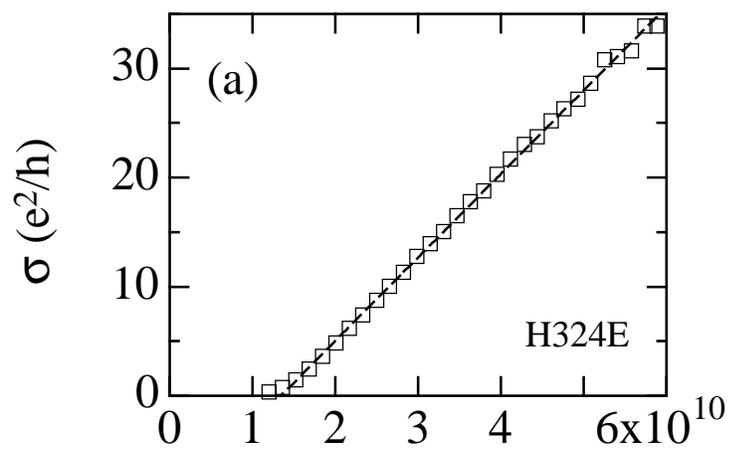
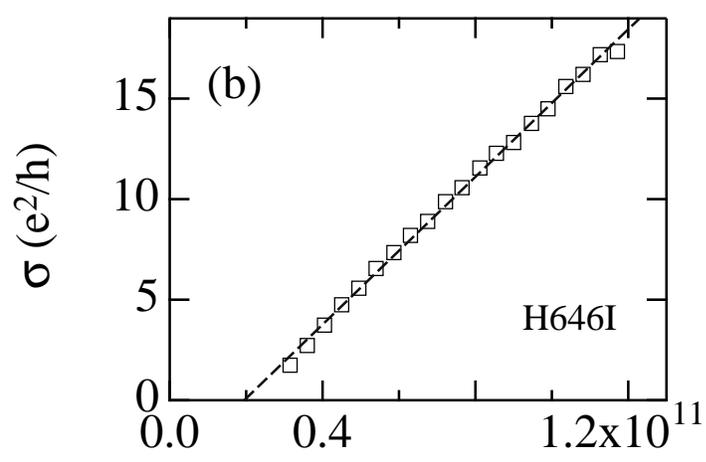
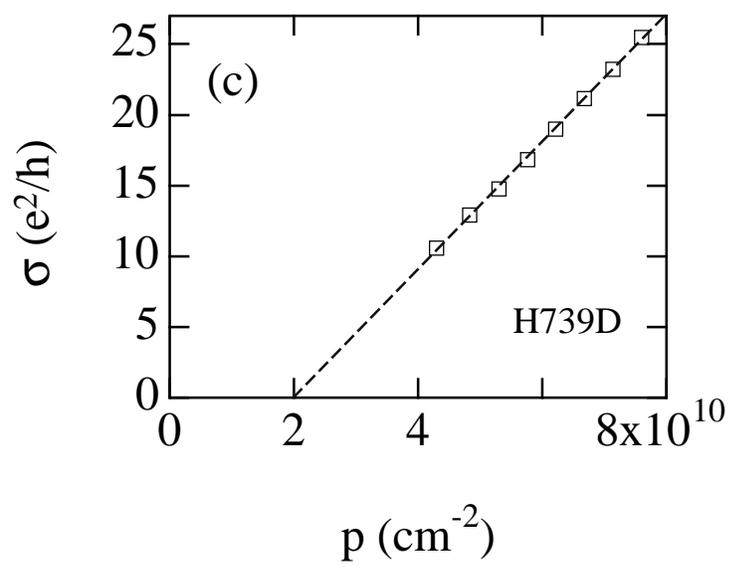
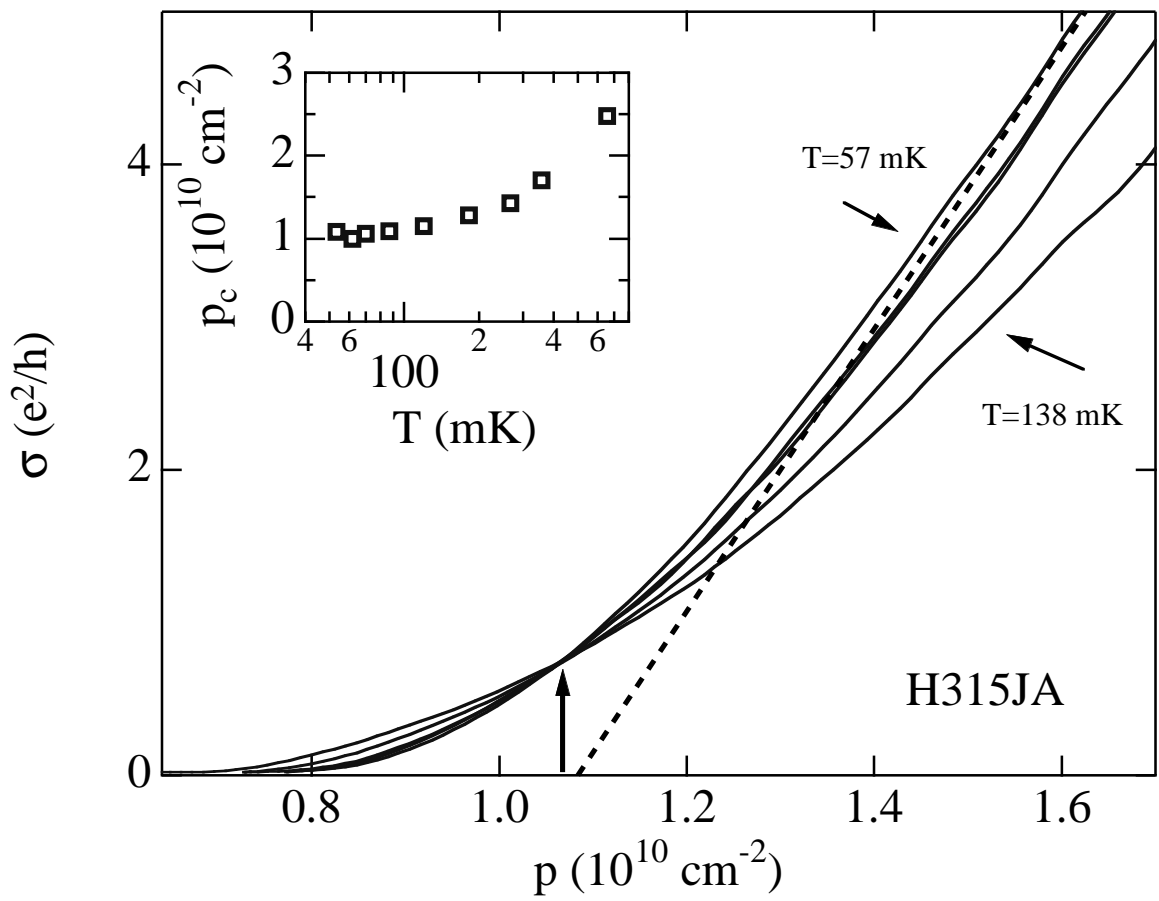
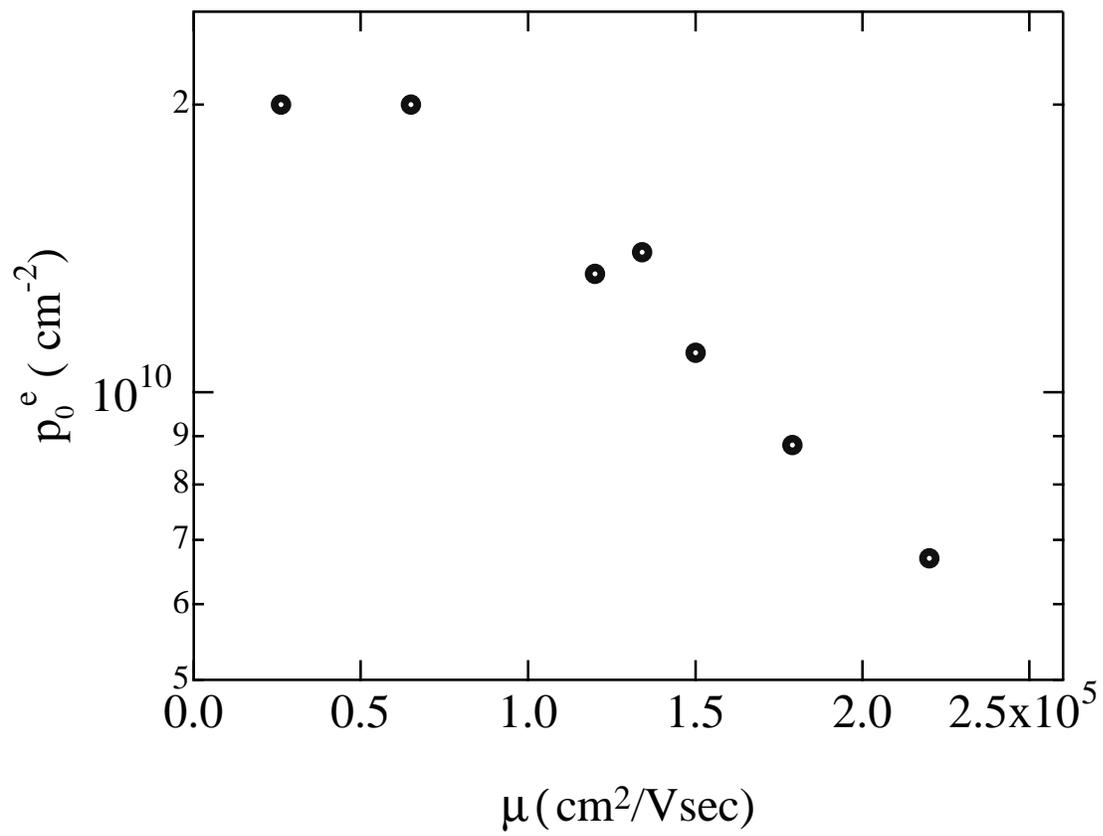
\end{document}